\input harvmac

\def\p{\partial}
\def\ap{\alpha'}

\Title{EFI-96-49}{Strings from IIB Matrices}
\centerline{Miao Li}
\centerline{\it Enrico Fermi Institute}
\centerline{\it University of Chicago}
\centerline{\it 5640 Ellis Avenue, Chicago, IL 60637} 
\centerline{\tt mli@curie.uchicago.edu}

\bigskip
\centerline{\it }
\centerline{\it }
\centerline{\it } 
\bigskip

D-string action is constructed from IIB matrices,
a spacetime commutator is essential in this construction. This
hints at the central role of the spacetime uncertainty relation
in a unified formulation of strings. Vertex operators of fundamental
strings are also discussed.

\Date{December 1996} 

\nref\bfss{T. Banks, W. Fischler, S. H. Shenker and L. Susskind,
hep-th/9610043.}
\nref\ikkt{N. Ishibashi, H. Kawai, Y. Kitazawa and A. Tsuchiya,
hep-th/9612115.}
\nref\bd{M. Berkooz and M. R. Douglas, hep-th/9610236.}
\nref\vp{V. Periwal, hep-th/9611132.}
\nref\sgrt{L. Susskind, hep-th/9611164; O. Ganor, S. Ramgoolam and
W. Taylor IV, hep-th/9611202.}
\nref\ab{O. Aharony and M. Berkooz, hep-th/9611215; G. Lifschytz and
S. Mathur, hep-th/9612087.}
\nref\mike{M. R. Douglas, hep-th/9612126.}
\nref\miao{M. Li, hep-th/9612144.}
\nref\bss{T. Banks, N. Seiberg and S. H. Shenker, hep-th/9612157.}
\nref\ks{S. Kachru and E. Silverstein, hep-th/9612162.}
\nref\tamy{T. Yoneya, in ``Wandering in the Fields", eds. K. Kawarabayashi and
A. Ukawa (World Scientific, 1987), p. 419: see also
{\it String Theory and Quantum
Gravity} in ``Quantum String Theory", eds. N. Kawamoto 
and T. Kugo (Springer, 1988), p. 23;
T. Yoneya, Mod. Phys. Lett. {\bf A4}, 1587(1989).}
\nref\ly{M. Li and T. Yoneya, hep-th/9611072.}
\nref\ml{M. Li, to appear.}
\nref\dhn{B. de Wit, J. Hoppe and H. Nicolai, Nucl. Phys. B305 [FS 23]
(1988) 545.}

\newsec{Introduction}

The recently proposed matrix model of M-theory \bfss\ may represent 
the beginning of our understanding of the new geometry required 
for a complete formulation of strings and branes. A number of issues 
in the BFSS matrix model have been studied
in \bd-\ks. Although there are problems, this matrix model has successfully
passed several consistency tests. Another matrix model, modeling the
IIB strings, has also been constructed \ikkt. There is no Hamiltonian
for this matrix model. There are some advantages to
work with an action principle in which both space and time are dynamically
generated. Lorentz invariance is manifest, also problems hard to treat
with the IMF Hamiltonian can be addressed. 

Here we shall study a couple of simple problems in the IKKT matrix model.
First, we wish to see how the action of multiple D-strings can be
recovered in this model. In doing so, we hit upon a highly suggestive
commutation relation concerning space and time, whose physics is discussed
in \tamy\ and \ly. Second, we discuss the relation between the operators of
the Wilson line type and the vertex operators of the fundamental strings. 
We point out that something crucial is missing in this strategy.

\newsec{D-string Action}

We start with the action proposed in \ikkt:

\eqn\red{S={1\over 2\pi (\ap)^2g_s}\Tr\left( {1\over 4}[X^\mu ,X^\nu]^2
+{1\over 2}\bar{\psi}\gamma^\mu [X_\mu ,\psi ]\right),}
where we have dropped the chemical potential term proportional 
to $\Tr 1$, which
is important to obtain a Born-Infeld action. We also corrected the
sign error in \ikkt\ (it is possible that the authors of \ikkt\
work with the Euclidean signature). Also the normalization of the action
is different from that of \ikkt. This normalization is justified by
the correct D-string action derived in this section, and also by
consideration of toroidal compactifications \ml.
The above action plus the 
chemical potential term, as 
explained in \ikkt, can be viewed as an effective action derived
from the large N reduced 10 dimensional ${\cal N}=1$ super Yang-Mills.
Therefore, there are two ways to interpret this action. One is that
of the effective action, so everything derived from it is not to be
treated quantum mechanically. Another is that it is just the large
N reduced model, so one also considers quantum fluctuation with this
action directly. We shall adopt both view-points here, whichever
is appropriate in the context of the special problem being concerned.

A large N solution satisfying $[X^\mu,[X^\nu,X_\mu]]=0$, adapted from
the ansatz for a transverse membrane in the BFSS matrix model, is
interpreted as a D-string configuration in \ikkt. The interaction properties
between two D-strings are calculated using \red\ and shown to agree
with what expected \ikkt. The ansatz is
\eqn\ans{\eqalign{X_0&=-Tp, \quad X_1=Lq, \cr
X_i &=0, \quad i\ge 2,\quad  [q,p]=i{2\pi\over N},}}
where the spectrum of $q$ and $p$ is $(0, 2\pi)$.
We shall use the following identification of the world-sheet coordinates
with $p$ and $q$: $\sigma =Tp$, $\tau=Lq$. It appears strange that
we have identified the world-sheet space with $X_0$ and the 
world-sheet time with $X_1$. The reason will become clear later.
With this identification, the world-sheet coordinates become noncommutative,
and unlike $[q,p]$ which tends to $0$ in the large N limit, the commutator
\eqn\comm{[\tau, \sigma ]=i{2\pi LT\over N}=2\pi i\ap}
is always finite,
where we have used the result $LT/N=\ap$ obtained in \ikkt. 
Thus, the space and time viewed on the D-string world-sheet become
noncommutative. The commutator \comm\ is precisely the form needed
to formulate the spacetime uncertainty relation advocated in \tamy\
\ly\ in a more rigorous fashion. From \comm, one may write
$X_0=2\pi i\ap\p_\tau$, $X_1=2\pi i\ap\p_\sigma$.

For a single D-string, any large $N$ matrix can be viewed as a function
of $p$ and $q$, or as a function of $\tau$ and $\sigma$. Thus, a matrix
appearing in the action \red\ becomes a world-sheet field. In the large
world-sheet limit, the range of both $\tau$ and $\sigma$ is infinite, and
a N large matrix such as $X^i$ can be written as
\eqn\fouri{X^i=\int d^2k X^i(k)e^{i(k_0\tau +k_1\sigma)},}
this is the continuous analogue of the discrete Fourier series on a 
quantum torus \bfss. A commutator such as $[A, B]$ is now replaced
by
\eqn\rule{[A,B]=2\pi i\ap\{A,B\}}
with the Poisson bracket $\{A,B\}=\p_\tau A\p_\sigma B -
\p_\sigma A\p_\tau B$. A trace is replaced by
$$\Tr\rightarrow \int {d^2\sigma\over (2\pi)^2 \ap}.$$

To reproduce the D-string action for a single string, we use the
following identification between matrices and fields
\eqn\iden{\eqalign{X_0 &=-\sigma +2\pi\ap A_0(\tau,\sigma), \quad
X_1=\tau +2\pi \ap A_1(\tau,\sigma), \cr
X_i &=X_i(\tau,\sigma),\quad \psi =\psi (\tau,\sigma).}}
Use the rule \rule, it is straightforward to calculate commutators:
$$\eqalign{ [X_0,X_1]&=2\pi i\alpha' (1+2\pi \ap F_{01}), 
\quad [X_\alpha, X_i]
=2\pi i\ap D_\alpha X_i,\cr
[X_\alpha, \psi ]&=2\pi i\ap D_\alpha \psi,}$$
where $F_{01}=\p_0A_1-\p_1A_0+2\pi\ap \{A_0, A_1\}$ and $D_\alpha
A=\p_\alpha A +2\pi\ap \{A_\alpha, A\}$. We see the $U(1)$ field
strength comes out right. This is one of the reasons to use the
identification \iden. Of course, to reproduce the field strength,
it is eqaully good to use the identification $X_0=-\tau +2\pi\ap A_1$
and $X_1=\sigma +2\pi\ap A_0$, so that the world-sheet time is
identified with $X_0$ and the world-sheet space is identified with
$X_1$. But with this scheme, the time derivative terms of other fields
will have the wrong sign in the action derived from \red.

One remarkable feature of the above discussion is that what we have
obtained is not a usual $U(1)$ gauge theory. The field stength
$F_{01}$ receives correction at the order $O(\ap)$, and the covariant
derivatives also receive corrections at the same order. This is due to
the fact that we have started with a large N gauge theory. The gauge
transformation, for instance $\delta X_0 =i[\epsilon, X_0]$
becomes $\delta A_0=\p_0\epsilon +2\pi\ap \{A_0, \epsilon\}$. This is
just the remanent of the Schild action as a starting point in the
derivation of \red\ \ikkt. As we shall see, in a more careful treatment,
there are more higher order corrections in $\ap$. It is straightforward
to write down the D-string action now. We just need to take a look at the
bosonic part
\eqn\ds{S_B={1\over 2\pi\ap g_s}\int d^2\sigma{1\over 2}\left(
1+(2\pi\ap)^2F_{01}^2-D_\alpha X^iD^\alpha X^i\right).}
This is the correct action, at least in the leading order in $\ap$.
The coefficient of the term $F^2$ matches the standard one.

We turn to the case of multiple D-strings. The N parallel D-string
configuration is given by
$$\eqalign{X_0&=(-\sigma \delta_{ab}), \quad X_1=(\tau\delta_{ab}),\cr
X^i&=\hbox{diag} (X^i_a),}$$
where we have used the block diagonalized matrices, so $a, b=1,
\dots, N$, each $X^i_a$ is a large N matrix and is proportional to
the large N identity matrix. In this spirit, we define $N\times N$ 
matrix-valued fields
\eqn\nona{\eqalign{X_0&=(-\sigma \delta_{ab} +2\pi\ap A_0^{ab}(\tau,\sigma)),
\quad
\quad X_1=(\tau\delta_{ab}+2\pi\ap A_1^{ab}(\tau,\sigma)),\cr
X_i &=(X_i^{ab}(\tau,\sigma)),\quad \psi =(\psi^{ab}(\tau,\sigma)).}}

In computing commutators of matrices, there are two types of product
to deal with. One type is associated to the matrix product with indices
$a,b$, this will give rise to the $U(N)$ gauge theory. Another type 
is the product of entries of the matrices, each is by itself a large N
matrix parametrized as a function of $\tau$ and $\sigma$. As the rule
\rule\ shows, this product is noncommutative too. As we shall see later,
this noncommutativeness will introduce terms in higher powers of $\ap$.
For the moment, we consider only the leading order. Thus
\eqn\field{\eqalign{[X_0,X_1]&=2\pi i\ap (1+2\pi\ap F_{01}), \quad
[X_\alpha, X^i]=2\pi i\ap D_\alpha X^i,\cr
[X^i,X^j]&=[X^i(\tau,\sigma), X^j(\tau,\sigma)],\quad
[X_\alpha,\psi]=2\pi i\ap D_\alpha\psi,}}
where the field strength and the covariant derivatives are the
standard ones for the $U(N)$ gauge theory. Subsituting the above result
with the rule $\Tr=\int d^2\sigma/(4\pi^2\ap )\tr$, $\tr$ is the trace
taken for $N\times N$ matrices, the 1+1 dimensional ${\cal N}=8$ super
Yang-Mills theory is obtained.

Now, consider $\ap$ corrections. These are sourses for the unusual gauge
theory for a single D-string discussed above. To systemacally compute
these corrections, we need to introduce some ``normal representation''
for a large N matrix. Let $f$ a large N matrix as a function of
$\tau$ and $\sigma$. We introduce its normal representation as
\eqn\norm{f(\tau,\sigma)=\int d^2\sigma f(k)e^{i(k_0\tau +
k_1\sigma)}.}
Apparently, due to the commutator \comm, $fg\ne gf$. The commutator
in the leading order in $\ap$ is given by \rule. In order to
distinquish between the large N matrix product and the usual product
of functions, we introduce $f*g$ to denote the large N matrix
product. By virtue of the formula $e^A*e^B=e^{1/2[A,B]}e^{A+B}$ valid
for a constant $[A,B]$,
$$e^{i(k_0\tau+k_1\sigma)} *e^{i(k'_0\tau+k'_1\sigma)}
=e^{i\ap\pi (k_1k'_0-k_0k'_1)}e^{i[(k_0+k'_0)\tau+(k_1+k'_1)\sigma]}.$$
Substituting the above relation to $f*g$ and using the normal representation
\norm, we find
\eqn\corr{f*g(\tau,\sigma)=fg(\tau,\sigma)+\sum_{n\ge 1}
(i\ap\pi )^n\{f,g\}_n,}
where we introduced the higher Poisson brackets
$$\{f,g\}_n={1\over n!}\left(\p^1_\tau\p^2_\sigma - \p^1_\sigma
\p^2_\tau\right)^nfg,$$
where $\p^1_\tau$ is $\p_\tau$ acting only on the first function, etc.

From \corr\ we see that the first term in the commutator $f*g-g*f$
is just what given in \rule. There are infinitely many more terms in this
commutator, all in odd powers of $\ap$. Also, there are corrections to
a $N\times N$ matrix commutator in the multiple D-string theory. For
instance
\eqn\corre{X^i*X^j-X^j*X^i=[X^i,X^j]+\sum_{n\ge 1}(i\ap\pi )^n
[X^i, X^j]_n,}
where
$$[X^i, X^j ]_n={1\over n!}\left(\p^1_\tau\p^2_\sigma - \p^1_\sigma
\p^2_\tau\right)^n[X^i,X^j].$$
To compute $(X^i*X^j-X^j*X^i)*(X^i*X^j-X^j*X^i)$, just apply \corr\
one more time. It is not necessary to do this in the action, since
the correction terms are total derivatives. This reflects the fact
that in taking trace, $\Tr A*B=\Tr B*A$.

In defining the normal representation \norm, one may replace 
$\exp(i(k_0\tau +k_1\sigma))$ by, say $\exp(ik_0\tau)\exp(ik_1\sigma)$.
This amounts to a redefinition of field $f$. The resulting commutators
such as the one in \corre\ will not respect the world-sheet Lorentz
invariance. We also should mention that high order corrections exist
in the BFSS matrix model when a certain configuration is considered.
However, the Planck constant there is $2\pi/N$ instead of $\ap$, so
higher orders are supressed in the large N limit. It is possible that
for certain problems, higher orders in that context will become important.

To summarize, the action of the multiple D-strings is reproduced from
the IIB matrix model, with high order corrections in $\ap$. The scheme
used here to construct the effective D-string action as well as a similar
scheme used to construct brane actions in the matrix model of M theory
\bss\ reproduces only the lowest mass states. Excited stringy states are 
absent. This may be due to the fact that we have assumed that the 
fluctutations such as those in \nona\ are more or less smooth functions 
of $\tau$ and $\sigma$, so highly excited open string states are to be 
found in singular fluctuations.

It also remains
to incoporate the Dirac-Born-Infeld action into this scheme, in which
there are high order corrections too. We shall discuss this in the next 
section. Of course, it should be interesting to check 
these corrections against direct string calculations.
The most suggestive of the derivation in this section
is the commutator \comm, which should be a universal feature in
a complete formulation of string theory. This commutator may form
the mathematical foundation for the spacetime uncertainty relation
discussed in \tamy\ \ly.

\newsec{Born-Infeld Action}

To derive the Born-Infeld action, we will work with a small area cell
$\Delta^2 \sigma$ of the world-sheet. We will also work with the 
Euclidean signature. The reason for this is that the Lagrangian
for a single D-string as written in \ds\ is not equal to
$$\det(\eta_{\alpha\beta}+\p_\alpha X^i\p_\beta X^i +
2\pi\ap F_{\alpha\beta}),$$
since the constant term has the wrong sign. We will see that rewriting
this part as a determinant is an important step in deriving the
Born-Infeld action.

We assume that the area element $\Delta^2\sigma$ is built from $N$
D-instantons. Following \ikkt, we shall let $N$ be a variable, so 
the ratio $\Delta^2\sigma/(4\pi^2 N)=\tilde{\alpha}$ is not equated to 
$\ap$. In this area element, we have $[\tau,\sigma]=2\pi i\tilde{\alpha}$,
and the trace $\Tr\rightarrow \Delta^2\sigma/(4\pi^2\tilde{\alpha})$.
The Eucliean action serving as our starting point is
\eqn\eucl{S=-{1\over 2\pi (\ap)^2g_s}\Tr\left({1\over 4}[X^\mu,X^\nu]^2
+{1\over 2}\bar{\psi}\gamma^\mu [X_\mu,\psi]\right)+{\pi\over g_s}
\Tr 1.}
We still use the ansatz \iden\ to define fluctutations on the D-string.
The relevant commutators are
$$\eqalign{[X_0,X_1]&=2\pi i\tilde{\alpha}(1+2\pi\ap F_{01}),\quad
[X_\alpha,X_i]=2\pi i\tilde{\alpha}D_\alpha X_i,\cr
[X_\alpha , \psi]&=2\pi i\tilde{\alpha}D_\alpha\psi,}$$
we ignored other terms with higher powers of $\tilde{\alpha}$.
Substituting the above result into the action \eucl, we find
\eqn\vari{{\Delta^2\sigma\over 4\pi (\ap)^2g_s}\left(\tilde{\alpha}
(1+(2\pi\ap)^2F_{01}+D_\alpha X_i D_\alpha X_i)-{i\over 2\pi}
\bar{\psi}\gamma^\alpha
D_\alpha\psi +(\ap)^2/\tilde{\alpha}\right).}
Note that the fermionic term is independent of $\tilde{\alpha}$, and
this is important in order to preserve SUSY in the resulting 
Born-Infeld action. The term linear in $\tilde{\alpha}$ is just
the following determinant
\eqn\detm{\det =\det(\delta_{\alpha\beta}+\p_\alpha X^i\p_\beta X^i +
2\pi\ap F_{\alpha\beta}).}

Taking variation of \vari\ with respect to $\tilde{\alpha}$ determines
$$\tilde{\alpha}={\ap\over \sqrt{\det}}.$$
Thus, we obtain the Born-Infeld action
\eqn\bia{S={\Delta^2\sigma\over 2\pi\ap g_s}\left(\sqrt{\det}
-{i\over 4\pi\ap}\bar{\psi}\gamma^\alpha D_\alpha\psi\right).}
This action is invaraint under the supersymmetry transformtion
\eqn\susyt{\eqalign{\delta A_\alpha &=\bar{\psi}\gamma_\alpha\epsilon,
\quad \delta X_i=\bar{\psi}\gamma_i\epsilon,\cr
\delta\psi &={2\pi i\ap\over \sqrt{\det}}\left({1\over 2}\gamma^{\alpha\beta}
F_{\alpha\beta}+\gamma^\alpha\gamma^iD_\alpha X_i\right)\epsilon.}}
Of course, both the action and the SUSY transformation make more sense
in the Minskowski space.

$\tilde{\alpha}$ is equal to $\ap$ when the world-sheet
fluctutaion is small, in which case $\sqrt{\det}\sim 1$. 
$(\tilde{\alpha})^{-1}$ is the local density of D-istantons. Since
the area element is corrected by a factor $\sqrt{\det}$, our formula
for $\tilde{\alpha}$ says that the real D-instanton density is always
a constant $1/\ap$. The Born-Infeld
action contains high order $\ap$ corrections. As we discussed in the
previous section, the noncommutativeness of the world-sheet coordinates
also introduces high order $\ap$ corrections. To treat all these 
corrections in a uniform manner, one has to replace $\ap$ in formulas
such as \corr\ by $\tilde{\alpha}$, and then take variation of the
whole action with respect to  $\tilde{\alpha}$. Although it is tedious
to carry out this procedure, the method is rather systematic.

\newsec{Vertex Operators of Fundamental Strings}

The suggestion in this section is highly conjectural.

It is suggested in \ikkt\ that the Wilson line operator
\eqn\wil{W(p(\sigma))=\Tr P \exp(i\int p_\mu (\sigma)X^\mu d\sigma)}
should be interpreted as the creation operator of a closed string state
with momentum density $p(\sigma)$, and some preliminary evidence in
supporting this interpretation is provided there. It is well-known
that the wave functions are directly related to vertex operators of
strings in computing scattering amplitudes. A general correlation function 
in the matrix model is then
\eqn\corrl{\langle W(p_1)W(p_2)\cdots W(p_n)\rangle.}
It is natural to conjecture that this correlation function is the one
of $n$ off-shell string states, containing all loops weighted by 
$g_s$. Curiously, the appearance of the string coupling in the matrix model 
action \red\ is not in $g_s^2$, but in $g_s$. The origin of this feature is
that the authors of \ikkt\ intended to reproduce D-string interaction
in which open string states are the relevant quanta. It is interesting
to see how the usual closed string coupling $g_s^2$ comes out correctly
in the correlation function \corrl.

Here we shall attempt to extract vertex operators from the Wilson line
operator \wil\ and similar operators. The first problem we need to
solve is to compute the trace in the Wilson lines. To represent
a large N matrix, one can either appeal to the spherical basis or
to the toroidal basis \dhn. We believe that different bases correspond
to contribution of different loops. Of course perturbative expansion
of the action \red\ also contains different loops. The correlation
function at a given loop therefore receives contribution not only
from different representaion of the large N matrices, but also from
perturbative expansion of the action. For simplicity, we use the
toroidal representation. Introduce $p$ and $q$ with $[q,p]=i\hbar =
i2\pi /N$. In the large N limit, matrices become functions of $p$
and $q$. Use $(q^1,q^2)=(q,p)$, then $[q^\alpha, q^\beta]=i\hbar
\epsilon^{\alpha\beta}$. In the large N limit, the Wilson line
in the leading order can be written as
\eqn\lead{W(p)=\int {d^2q\over 2\pi\hbar}e^{iP_\mu X^\mu(q,p)},}
where $P_\mu =\int p_\mu (\sigma)d\sigma$.
So this leading order term can be interpreted as the tachyon operator.
The correlation function \corrl\ is then interpreted as the correlation
of $n$ tachyon states starting at the one-loop, since we use
the toroidal representation here. The difficulty for this correspondence
to be exact lies in the fact that the matrix model action \red\ is
like the action of string field theory, we do not know how the world-sheet
action of the fundamental strings can be derived from it.
For instance, to compute the one-loop correlation function, the action
\red\ contributes only at the $O(g_s^0)$ level
in the present consideration, since we have already
used the toroidal representation for the Wilson line operators.
Only classical solution representing the vacuum contribute. This solution
is just the Goldstone modes $X^\mu=x^\mu 1_N$. 

Depite the above difficulty, we still want to push our straitegy to
its limit to see how much we can get out of it. Since $q^\alpha$ are
noncommutative, the Wilson line operator \wil\ contains infinitely
many terms in higher orders in $\hbar$. To compute these terms 
systematically, one may use the method introduced in the previous
section. One first uses the normal reprersentation for $X(p,q)$, then
arranges the Wilson line operator in a normal representation. Here 
we write down only the second term in the expansion in $\hbar$:
\eqn\expan{W(p)=\int {d^2q\over 2\pi\hbar}e^{iP_\mu X^\mu}
\left(1-i\hbar \epsilon^{\alpha\beta}
\xi_{\mu\nu} \p_\alpha X^\mu\p_\beta X^\nu +\cdots\right),}
where
$$\xi_{\mu\nu}={1\over 2}\int_{\sigma_1<\sigma_2}p_\mu(\sigma_1)
p_\nu(\sigma_2)d\sigma_1d\sigma_2.$$
(To compute the expansion \expan\ systematically, one needs the
formula 
$$A_1*A_2*\cdots *A_n=\exp\left({i\over 2}\hbar\sum_{i<j}\epsilon^{\alpha\beta}
\p^i_\alpha\p^j_\beta\right)A_1A_2\cdots A_n,$$
where $\p^i_\alpha$ is $\p_\alpha$ acting on the i-th factor.)
We see that the vertex operator of the anti-symmetric tensor is reproduced
as the second term. The polarization tensor is nonvanishing only when
the density $p_\mu(\sigma)$ is not constant. This sounds reasonable,
since this mode is an excited state above the tachyon.
One more difficulty with this approach appears here. It is impossible
to construct the graviton vertex operator from the Wilson line or
more complicated operator such as $\Tr[X^\mu,X^\nu]P\exp(i\int p_\lambda
(\sigma)X^\lambda d\sigma)$. There is one place we know graviton
exists. In computing
the interaction between two parallel static D-strings \ikkt, 
the effect of the bosonic fluctuations cancels that of the fermionic
fluctuations. In the closed string channel, this is interpreted as
the cancelation between the exchange of graviton and dilaton and
the exchange of the qaunta of R-R anti-symmetric tensor field.
It is interesting to complete the calculation of the Wilson line
tapole in the background in a D-string, as suggetsed in \ikkt, to
see whether the graviton vertex operator is contained in the Wilson
line. 

The vertex operator for the R-R tensor field can not be constructed
directly from matrices either. The natural candidate  $\Tr\bar{\psi}
\gamma^{\mu\nu}\psi P\exp(i\int p_\lambda (\sigma)X^\lambda d\sigma)$ 
vanishes, since $\psi$ is chiral. Just as in the usual world-sheet 
approach, only the
vertex for the field strength exists, which is $\Tr\bar{\psi}
\gamma^{\mu\nu}p_\sigma\gamma^\sigma\psi
P\exp(i\int p_\lambda (\sigma)X^\lambda d\sigma)$. Of course, it is
also necessary to generate both the left moving and the right moving
world-sheet spinors from a single $\psi$.

One expects from the definition of the Wilson line \wil\ that only 
a loop structure emerges, not a world-sheet as we have suggested.
This puzzle may be resolved by the following interpretation. Upon
large N expansion \expan, there are infinitely many vertex operators,
inserted at the puncture $(q,p)$. The whole effect is then to open
this puncture to a loop. So in a toroidal representation of the
large N matrices, a Wilson line really represents an open torus whose
boundary is a circle. In the spirit of string field theory, what one
expects from action are vertices represented by a Riemann surface
with multiple boundaries. The contractions
between inserted operators \corrl\ and the action, and the contractions
between operators contained in the action, will help to form a Riemann
surface of higher genus. In any case, it is worthwhile to pursue the
matter along
this line further, although it seems there are many difficulties.

\newsec{Discussion}

We have constructed the action of multiple D-strings from IIB matrices.
One nice feature of this construction is the commutation relation for the
D-string world-sheet coordinates, which reflect time and the space these
D-strings are embeded in. This relation implies high order $\ap$
corrections, in terms of the lowest mass states. The open problem is
to construct the excited open string states living on D-strings. 
The spacetime uncertainty relation implied
by the commutation relation was discussed some time ago \tamy\ and
has also received attention recently in the more general context
involving D-objects \ly. We believe that this uncertainty principle
plays a central role in our future understanding of the foundation of
string theory. And here it is the first time for us to see its mathematical
underpinning in the IIB matrix model, rather than by analysis of perturbative
amplitudes \tamy.

A suggestion to extract vertex operators of fundamental strings 
from operators of the Wilson line type is made. We are beginning to see 
difficulty in this scheme. For one thing, we do not know how to construct
a simple vertex operator such as that of the graviton. For another,
we are not yet able to recover a world-sheet
picture from the matrix model action \red. This difficulty is perhaps
of technical nature rather than of conceptual one. To conclude, we
are just beginning to see the tip of the huge iceberg burried in
the formal matrix model.

\noindent{\bf Acknowledgments} 

We thank Emil Martinec for comments and discussions on other matters
in the matrix models.
This work was supported by DOE grant DE-FG02-90ER-40560 and NSF grant
PHY 91-23780.

\listrefs

\end